\documentclass[twocolumn,showpacs,preprintnumbers,amsmath,amssymb]{revtex4}

\usepackage{graphicx}
\usepackage{dcolumn}
\usepackage{bm}
\begin{document}

\preprint{APS/123-QED}

\title{Deformed Special Relativity with an energy barrier of a minimum speed}
\author{Cl\'audio Nassif\\
        (e-mail: {\bf cnassif@cbpf.br})}
 \altaffiliation{{\bf CBPF}: Centro Brasileiro de Pesquisas F\'isicas, Rua Dr.Xavier Sigaud 150, 22290-180, Rio de Janeiro, Brazil.}

\begin{abstract}

 This paper aims to introduce a new principle of symmetry in the flat space-time by means of the elimination of the classical idea
 of rest, and by including a universal minimum limit of speed in the quantum world. Such a limit, unattainable by the particles,
 represents a preferred inertial reference frame associated with a universal background field that breaks Lorentz symmetry.
 So there emerges a new relativistic dynamics where a minimum speed forms an inferior energy barrier. One of the interesting
 implications of the existence of such a minimum speed is that it prevents the absolute zero temperature for an ultracold gas according
 to the third law of thermodynamics. So we will be able to provide a fundamental dynamical explanation for the third law by means
 of a connection between such a phenomenological law and the new relativistic dynamics with a minimum speed.

 DOI: 10.1142/S021827181001652X
 
\end{abstract}

\maketitle

\section{\label{sec:level1} Introduction}

 In 1905, in the most meaningful article entitled {\it ``On the Electrodynamics of Moving Bodies"}, Einstein solved the old incompatibility
 between classical
 mechanics and Maxwell theory, leading to a reformulation of our conception of space and time. In order to do that, he tried to preserve
 the symmetries of
Maxwell equations by postulating the speed of light ($c$) as invariant under any change of reference frame. At the end of his life, he continued searching in
vain for the beauty of new symmetries in order to unify gravitation with electromagnetism, from where there would emerge a more fundamental explanation
for the quantum phenomena by means of a theory of quantum gravity.

 Still inspired by the seductive search for new fundamental symmetries in Nature\cite{1}, the present article attempts to implement a
 uniform
background field into the  flat space-time. Such a background field connected to a uniform vacuum energy density represents a preferred reference frame,
 which leads us to postulate a universal and invariant minimum limit of speed for particles with very large wavelengths (very low
 energies).

 The idea that some symmetries of a fundamental theory of quantum gravity may have non trivial consequences for cosmology and particle
 physics at very
low energies is interesting and indeed quite reasonable. So it seems that the idea of a universal minimum speed as one of the first attempts of
Lorentz symmetry violation could have the origin from a fundamental theory of quantum gravity at very low energies (very large wavelengths).

 Besides quantum gravity for the Planck minimum length $l_P$ (very high energies), the new symmetry idea of a minimum speed $V$ could
 appear due to the
indispensable presence of gravity at quantum level for particles with very large wavelengths (very low energies). So we expect that such a universal
minimum speed $V$ also depends on fundamental constants as for instance $G$ (gravitation) and $\hbar$ (quantum mechanics)\cite{2}. In
 this sense, there could
be a relation between $V$ and $l_P$ since $l_P\propto (G\hbar)^{1/2}$. The origin of $V$ and a possible connection between $V$ and $l_P$ shall be deeply investigated in a further work\cite{2}.

 The hypothesis of the lowest non-null limit of speed for low energies ($v<<c$) in the space-time results in the following physical
 reasoning:

- In non-relativistic quantum mechanics, the plane wave wave-function ($Ae^{\pm ipx/\hbar}$) which represents a free particle is an idealisation that
is impossible to conceive under
physical reality. In the event of such an idealized plane wave, it would be possible to find with certainty the reference frame that cancels its momentum
 ($p=0$),
so that the uncertainty on its position would be $\Delta x=\infty$. However, the presence of an unattainable minimum limit of speed emerges in order to prevent
 the
ideal case of a plane wave wave-function ($p=constant$ or $\Delta p=0$). This means that there is no perfect inertial motion ($v=constant$) such as a plane
 wave,
except the privileged reference frame of a universal background field connected to an unattainable minimum limit of speed $V$, where $p$ would vanish.
 However,
since such a minimum speed $V$ (universal background frame) is unattainable for the particles with low energies (large length scales), their momentum can
actually never vanish when one tries to be closer to such a preferred frame ($V$). On the other hand, according to Special Relativity (SR), the momentum
cannot be infinite since the maximum speed $c$ is also unattainable for a massive particle, except the photon ($v=c$) as it is a massless particle.

  This reasoning allows us to think that the photon ($v=c$) as well as the massive particles ($v<c$) are in equal-footing in the sense that
 it is not possible
 to find a reference frame at rest ($v_{relative}=0$) for any speed transformation in a space-time with both maximum and minimum speed
limits. Therefore, such a deformed special relativity was termed as Symmetrical Special Relativity (SSR)\cite{3}.

 The dynamics of particles in the presence of a universal background reference frame connected to $V$ is within a context of the ideas of
 Sciama\cite{4},
 Schr\"{o}dinger\cite{5} and Mach\cite{6}, where there should be an ``absolute" inertial reference frame in relation to which we have the
 inertia of all
 moving bodies. However, we must emphasize that the concept used here is not classical as machian ideas, since the lowest (unattainable)
 limit of speed $V$
 plays the role of a privileged (inertial) reference frame of a universal background field instead of the ``inertial" frame of fixed stars.

 It is interesting to notice that the idea of universal background field was sought in vain by Einstein\cite{7}, motivated firstly by
 Lorentz\cite{8}. It was Einstein who coined the term {\it ultra-referential} as the fundamental aspect of Reality to represent a universal
 background field\cite{9}\cite{10}\cite{11}\cite{12}\cite{13}. Based on such a concept, let us call {\it ultra-referential} $S_V$ to be the
 universal background field of a fundamental inertial reference frame connected to $V$ (see reference\cite{3}).

 The present theory (SSR) is a kind of deformed special relativity (DSR) with two invariant scales, namely the speed of light $c$ and a
 minimum speed $V$. DSR was first proposed by Camelia\cite{14}\cite{15}\cite{16}\cite{17}. It
 contains two invariant scales: speed of light $c$ and a minimum length scale (Planck length $l_P$ of quantum gravity). An alternate
 approach to DSR theory, inspired by that of Camelia, was proposed later by Smolin and Magueijo\cite{18}\cite{19}\cite{20}.

 Another extension of Special Relativity (SR) is known as triply special relativity, which is characterized by three invariant scales,
 namely the speed of light $c$, a mass $k$ and a length $R$\cite{21}. Still another generalization of SR is the quantizing of
 speeds\cite{22}, where Barrett-Crane spin foam model for quantum gravity with
positive cosmological constant was considered, encouraging the authors to look for a discrete spectrum of velocities and the physical
 implications of this effect, namely an effective deformed Poincar\'e symmetry.

 In a more recent paper\cite{3}, it was shown that the existence of a minimum (non-zero) speed $V$ connected to a background field with
 minimum energy leads to a tiny positive cosmological constant, being in agreement with observations ($\Lambda\sim 10^{-35}s^{-2}$). In
 fact, such an original physical result\cite{3} also encourage us to search for a deformed Poincar\'e symmetry with the presence of an
 invariant minimum speed $V$.

\section{\label{sec:level1}Transformations of space-time coordinates in the presence of the ultra-referential $S_V$}

 The classical notion we have about the inertial (galilean) reference frames, where the system at rest exists, is eliminated in SSR where
 $v>V$ (see Fig.1). However, if we consider classical systems composed of macroscopic bodies, the minimum speed V is neglected ($V=0$)
 and so we can reach a vanishing velocity ($v=0$), i.e., in the classical approximation ($V\rightarrow 0$), the
 ultra-referential $S_V$ is eliminated and simply replaced by the galilean reference frame S connected to a system at rest.

  Symmetrical Special Relativity (SSR) should contain three postulates, namely:

 1) the constancy of the speed of light $c$.

 2) the non-equivalence (asymmetry) of the reference frames, i.e., we cannot exchange the speed $v$ (of $S^{\prime }$) for $-v$ (of
 $S_V$) by the inverse transformations, since we cannot find the rest for $S^{\prime}$ (see Fig.1). Such an asymmetry was explored by
 means of speed transformations of SSR in a previous paper\cite{3}.

 3) the covariance of the ultra-referential $S_V$ (background frame) connected to an unattainable minimum limit of speed $V$ (Fig.1).

 \subsection{$(1+1)D$ space-time in SSR}

Let us assume the reference frame $S^{\prime}$ with a speed $v$ in relation to the ultra-referential $S_V$ according to Fig. 1.

\begin{figure}
\includegraphics[scale=0.8]{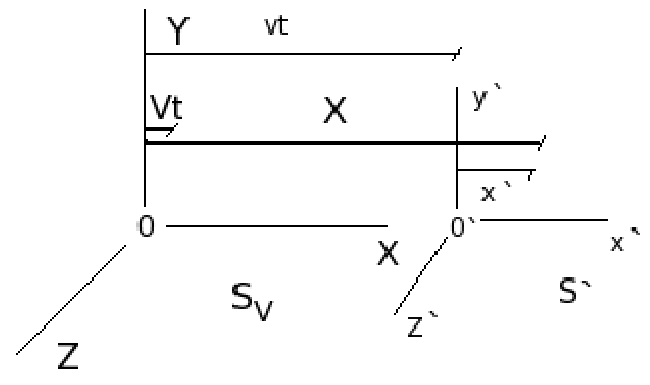}
\caption{$S^{\prime}$ moves with a velocity $v$ with respect to the background field of the covariant ultra-referential $S_V$.
 If $V\rightarrow 0$, $S_V$ is eliminated and thus the
 galilean frame $S$ takes place, recovering Lorentz transformations.}
\end{figure}

Hence, to simplify, consider the motion at only one spatial dimension, namely $(1+1)D$ space-time with background field $S_V$. So we write the following
transformations:

  \begin{equation}
 x^{\prime}=\Psi(X-\beta_{*}ct)=\Psi(X-vt+Vt),
  \end{equation}
where $\beta_{*}=\beta\epsilon=\beta(1-\alpha)$, being $\beta=v/c$ and
 $\alpha=V/v$, so that $\beta_{*}\rightarrow 0$ for $v\rightarrow V$ or $\alpha\rightarrow 1$.

 \begin{equation}
 t^{\prime}=\Psi\left(t-\frac{\beta_{*}X}{c}\right)=\Psi\left(t-\frac{vX}{c^2}+\frac{VX}{c^2}\right),
  \end{equation}
being $\vec v=v_x{\bf x}$, $\left|\vec v\right|=v_x=v$ and $v_*=\beta_*c=v-V$. $\left|\vec V\right|=V$, where $\vec V$ is a vector given in the
direction of $x$. In Fig.1, we consider, for instance, the motion to right in the direction of $x$ ($(1+1)D$ space). The $(3+1)D$ case will be explored in the next subsection.

  At first sight, $v_*$ can be negative, however as it will be shown in section 4, the limit $V$ forms an inferior energy barrier according
 to a new dynamical viewpoint of SSR, and so $v_*$ must be positive in physical reality.

 We have $\Psi=\frac{\sqrt{1-\alpha^2}}{\sqrt{1-\beta^2}}$ to be justified later. If $v<V$ ($v_*<0$ or $\alpha >1$), $\Psi$ would be
 imaginary, that is to say it is a non-physical factor. So we must have $v_*>0$ to be justified in section 4.

If we make $V\rightarrow 0$ ($\alpha\rightarrow 0$ or $v_*=v$), we recover Lorentz transformations, where the ultra-referential $S_V$ is eliminated and
 simply replaced by the galilean frame $S$ at rest for the classical observer.

 In order to get the transformations (1) and (2) above, let us consider the following more general transformations:
$x^{\prime}=\theta\gamma(X-\epsilon_1vt)$ and
 $t^{\prime}=\theta\gamma(t-\frac{\epsilon_2vX}{c^2})$, where $\theta$, $\epsilon_1$ and $\epsilon_2$ are factors (functions) to be determined. We hope all
 these
 factors depend on $\alpha$, such that, for $\alpha\rightarrow 0$ ($V\rightarrow
 0$), we recover Lorentz transformations as a particular case
 ($\theta=1$, $\epsilon_1=1$ and $\epsilon_2=1$). By using those transformations to perform
$[c^2t^{\prime 2}-x^{\prime 2}]$, we find the identity: $[c^2t^{\prime
 2}-x^{\prime 2}]=
\theta^2\gamma^2[c^2t^2-2\epsilon_1vtX+2\epsilon_2vtX-\epsilon_1^2v^2t^2+\frac{\epsilon_2^2v^2X^2}{c^2}-X^2]$.
 Since the metric tensor is diagonal, the crossed terms must vanish and so we assure that
$\epsilon_1=\epsilon_2=\epsilon$. Due to this fact, the crossed terms
 ($2\epsilon vtX$) are cancelled between themselves and finally we obtain $[c^2t^{\prime 2}-x^{\prime 2}]=
 \theta^2\gamma^2(1-\frac{\epsilon^2 v^2}{c^2})[c^2t^2-X^2]$. For
 $\alpha\rightarrow 0$ ($\epsilon=1$ and
$\theta=1$), we reinstate $[c^2t^{\prime 2}-x^{\prime 2}]=[c^2t^2-x^2]$ of SR. Now we write the following
transformations: $x^{\prime}=\theta\gamma(X-\epsilon
 vt)\equiv\theta\gamma(X-vt+\delta)$ and
$t^{\prime}=\theta\gamma(t-\frac{\epsilon
 vX}{c^2})\equiv\theta\gamma(t-\frac{vX}{c^2}+\Delta)$, where we assume $\delta=\delta(V)$ and $\Delta=\Delta(V)$, so that $\delta
 =\Delta=0$ for $V\rightarrow 0$, which implies $\epsilon=1$.
 So from such transformations we extract: $-vt+\delta(V)\equiv-\epsilon vt$ and
$-\frac{vX}{c^2}+\Delta(V)\equiv-\frac{\epsilon vX}{c^2}$, from where we obtain
 $\epsilon=(1-\frac{\delta(V)}{vt})=(1-\frac{c^2\Delta(V)}{vX})$. As
 $\epsilon$ is a dimensionless factor, we immediately conclude that $\delta(V)=Vt$ and
 $\Delta(V)=\frac{VX}{c^2}$, so that we find
$\epsilon=(1-\frac{V}{v})=(1-\alpha)$. On the other hand, we can determine $\theta$ as follows: $\theta$ is a function of $\alpha$ ($\theta(\alpha)$), such
 that
 $\theta=1$ for
 $\alpha=0$, which also leads to $\epsilon=1$ in order to recover Lorentz transformations. So, as $\epsilon$ depends on
 $\alpha$, we conclude that $\theta$ can also be expressed in terms of $\epsilon$, namely
 $\theta=\theta(\epsilon)=\theta[(1-\alpha)]$, where
$\epsilon=(1-\alpha)$. Therefore we can write
 $\theta=\theta[(1-\alpha)]=[f(\alpha)(1-\alpha)]^k$, where the exponent $k>0$. Such a positive value must be justified later
within a dynamical context (section 4).

 The function $f(\alpha)$ and $k$ will be estimated by satisfying the following conditions:

i) as $\theta=1$ for $\alpha=0$ ($V=0$), this implies $f(0)=1$.

ii) the function $\theta\gamma =
\frac{[f(\alpha)(1-\alpha)]^k}{(1-\beta^2)^{\frac{1}{2}}}=\frac{[f(\alpha)(1-\alpha)]^k}
{[(1+\beta)(1-\beta)]^{\frac{1}{2}}}$ should have a symmetrical behavior, that is to say it approaches to zero when closer to $V$ ($\alpha\rightarrow 1$), and
 in the same way to the infinite when closer to $c$ ($\beta\rightarrow 1$). In other words, this means that the numerator of the function $\theta\gamma$, which
 depends on $\alpha$ should have the same shape of its denominator, which depends on $\beta$. Due to such conditions, we naturally conclude that $k=1/2$ and
$f(\alpha)=(1+\alpha)$, so that $\theta\gamma =
\frac{[(1+\alpha)(1-\alpha)]^{\frac{1}{2}}}{[(1+\beta)(1-\beta)]^{\frac{1}{2}}}=
\frac{(1-\alpha^2)^{\frac{1}{2}}}{(1-\beta^2)^\frac{1}{2}}=\frac{\sqrt{1-V^2/v^2}}{\sqrt{1-v^2/c^2}}=\Psi$,
 where $\theta =(1-\alpha^2)^{1/2}=(1-V^2/v^2)^{1/2}$. In order justify the positive value of $k$ ($=1/2$), first of all
 we will study the dynamics of a particle submitted to a force in the same direction of its motion, so that the new relativistic power in
 SSR
 ($P_{ow}$) should be computed to show us that the minimum limit of speed $V$ works like an inferior energy barrier, namely
 $P_{ow}=vdp/dt$. So when we make
 such a derivative ($dp/dt$) of the new momentum $p=\Psi m_0v=\theta\gamma m_0v$ (eq.34), we are able to see an effective energy barrier of
 $V$, where a vacuum
 energy of the ultra-referential $S_V$ takes place, governing the dynamics of the massive particles (section 4).

The transformations shown in (1) and (2) are the direct transformations from $S_V$ [$X^{\mu}=(X,ict)$] to $S^{\prime}$
 [$x^{\prime\nu}=(x^{\prime},ict^{\prime})$], where we have $x^{\prime\nu}=\Omega^{\nu}_{\mu} X^{\mu}$ ($x^{\prime}=\Omega X$), so that we
 obtain the following matrix of transformation:

\begin{equation}
\displaystyle\Omega=
\begin{pmatrix}
\Psi & i\beta (1-\alpha)\Psi \\
-i\beta (1-\alpha)\Psi & \Psi
\end{pmatrix},
\end{equation}
such that $\Omega\rightarrow\ L$ (Lorentz matrix of rotation) for
 $\alpha\rightarrow 0$ ($\Psi\rightarrow\gamma$). We should investigate whether the transformations (3) form a group. However, such an
 investigation can form the basis of a further work.

We obtain $det\Omega
 =\frac{(1-\alpha^2)}{(1-\beta^2)}[1-\beta^2(1-\alpha)^2]$, where $0<det\Omega<1$. Since
$V$ ($S_V$) is unattainable ($v>V)$, this assures that $\alpha=V/v<1$
 and therefore the matrix $\Omega$
admits inverse ($det\Omega\neq 0$ $(>0)$). However $\Omega$ is a non-orthogonal matrix
($det\Omega\neq\pm 1$) and so it does not represent a rotation matrix
 ($det\Omega\neq 1$) in such a space-time due to the presence of the privileged frame of background field $S_V$ that breaks strongly the invariance of the
 norm of the 4-vector of SR (section 3). Actually such an effect ($det\Omega\approx 0$ for $\alpha\approx 1$) emerges from a new relativistic physics of SSR
 for treating much lower energies at ultra-infrared regime closer to $S_V$ (very large wavelengths).

 We notice that $det\Omega$ is a function of the speed $v$ with respect to $S_V$. In the approximation for $v>>V$ ($\alpha\approx 0$), we obtain
 $det\Omega\approx
 1$ and so we practically reinstate the rotational behavior of Lorentz matrix as a particular regime for higher energies. If we make $V\rightarrow 0$
 ($\alpha\rightarrow 0$), we exactly recover $det\Omega=1$.

The inverse transformations (from $S^{\prime}$ to $S_V$) are

 \begin{equation}
 X=\Psi^{\prime}(x^{\prime}+\beta_{*}ct^{\prime})=\Psi^{\prime}(x^{\prime}+vt^{\prime}-Vt^{\prime}),
  \end{equation}

 \begin{equation}
 t=\Psi^{\prime}\left(t^{\prime}+\frac{\beta_{*}
 x^{\prime}}{c}\right)=\Psi^{\prime}\left(t^{\prime}+\frac{vx^{\prime}}{c^2}-\frac{Vx^{\prime}}{c^2}\right).
  \end{equation}

In matrix form, we have the inverse transformation
 $X^{\mu}=\Omega^{\mu}_{\nu} x^{\prime\nu}$
 ($X=\Omega^{-1}x^{\prime}$), so that the inverse matrix is

\begin{equation}
\displaystyle\Omega^{-1}=
\begin{pmatrix}
\Psi^{\prime} & -i\beta (1-\alpha)\Psi^{\prime} \\
 i\beta (1-\alpha)\Psi^{\prime} & \Psi^{\prime}
\end{pmatrix},
\end{equation}
where we can show that $\Psi^{\prime}$=$\Psi^{-1}/[1-\beta^2(1-\alpha)^2]$, so that we must satisfy $\Omega^{-1}\Omega=I$.

 Indeed we have $\Psi^{\prime}\neq\Psi$ and therefore
 $\Omega^{-1}\neq\Omega^T$. This non-orthogonal aspect of
$\Omega$ has an important physical implication. In order to understand such an implication, let us first consider the orthogonal (e.g: rotation) aspect of
 Lorentz matrix in SR. Under SR, we have $\alpha=0$, so that
 $\Psi^{\prime}\rightarrow\gamma^{\prime}=\gamma=(1-\beta^2)^{-1/2}$.
  This symmetry ($\gamma^{\prime}=\gamma$, $L^{-1}=L^T$) happens because the galilean reference frames allow us to exchange the speed $v$ (of $S^{\prime}$)
 for $-v$ (of $S$) when we are at rest at
$S^{\prime}$. However, under SSR, since there is no rest at
 $S^{\prime}$, we cannot exchange $v$ (of $S^{\prime}$) for $-v$ (of $S_V$)
due to that asymmetry ($\Psi^{\prime}\neq\Psi$,
 $\Omega^{-1}\neq\Omega^T$). Due to this fact,
$S_V$ must be covariant, namely $V$ remains invariant for any change of reference frame in such a space-time. Thus we can notice that the paradox of twins,
 which appears due to the symmetry by exchange of $v$ for $-v$ in SR should be naturally eliminated in SSR, where only the reference frame $S^{\prime}$ can move
 with respect to $S_V$. So $S_V$ remains covariant (invariant for any change of reference frame).
  We have $det\Omega=\Psi^2[1-\beta^2(1-\alpha)^2]\Rightarrow
 [(det\Omega)\Psi^{-2}]=[1-\beta^2(1-\alpha)^2]$. So
we can alternatively write
 $\Psi^{\prime}$=$\Psi^{-1}/[1-\beta^2(1-\alpha)^2]=\Psi^{-1}/[(det\Omega)\Psi^{-2}]
=\Psi/det\Omega$. By inserting this result in (6) to replace
 $\Psi^{\prime}$, we obtain the relationship between the inverse matrix and the transposed matrix of $\Omega$, namely $\Omega^{-1}=\Omega^T/det\Omega$. Indeed
 $\Omega$ is a non-orthogonal matrix, since we have $det\Omega\neq\pm 1$.

  According to Fig.1, it is important to notice that a particle moving in one spatial dimension ($x$) goes only to right or to left, since the unattainable
 minimum limit of speed $V$, which represents the spatial aspect of the space-time in SSR, prevents it to stop ($v=0$) in the space. So it cannot return in the
 same spatial dimension $x$. On the other hand, in a complementary and symmetric way to $V$, the limit $c$, which represents the temporal aspect of the space-time,
  prevents to stop the marching of the time ($v_t=0$), and so avoiding to come back to the past (see eq.29). In short, we perceive that the basic ingredient of
 the space-time structure in SSR, namely the $(1+1)D$ space-time presents $x$ and $t$ in equal-footing in the sense that both of them are irreversible once the
 particle is moving to right or to left. Such an equal-footing ``$xt$" in SSR does not occurs in SR since we can stop the spatial motion
in SR ($v_x=0$) and so come back in $x$, but not in
 $t$. However, if we take into account more than one spatial dimension in SSR, at least two spatial dimensions ($xy$), thus the particle could return by moving in
 the additional (extra) dimension(s) $y$ ($z$). So SSR is able to provide the reason why we must have more than one (1) spatial dimension for representing movement
 in reality $(3+1)D$, although we could have one (1) spatial dimension just as a good practical approximation for some cases of classical space-time as in SR 
 (e.g.:a ball moving in a rectilinear path).

   The reasoning above leads us to conclude that the minimum limit $V$ has deep implications for understanding the irreversible aspect of
 the time connected to the spatial motion in $1D$. Such an irreversibility generated by SSR just for $(1+1)D$ ($xt$) space-time really
 deserves a deeper treatment in a future research.

\subsection{$(3+1)D$ space-time in SSR}

 In order to obtain general transformations for the case of $(3+1)D$ space-time, we should replace the one dimensional coordinates $X$ and
 $x^{\prime}$ by the $3$-vectors $\vec r$ and $\vec r^{\prime}$ so that we write: $\vec r=\vec r_{||}+\vec r_{T}$, being $\vec r_{||}$
given in the direction of the motion and $\vec r_{T}$ is given in the transversal direction of the motion $\vec v$.

 At the frame $S^{\prime}$ with velocity $\vec v$, we have the vector $\vec r^{\prime}$, where we generally write: $\vec r^{\prime}=\vec
 r^{\prime}_{||}+\vec r^{\prime}_{T}$.

In the classical $(3+1)D$ space-time of SR, of course we should have $\vec r^{\prime}_{T}=\vec r_{T}$ since there is no boost
 for the transversal direction, so that the modulus of $\vec r_{T}$ is always preserved for any reference frame. However, for $(3+1)D$
 space-time of SSR, there
 should be a transformation of the transversal vector such that $\vec r^{\prime}_{T}\neq\vec r_{T}$. So let us admit the following
transformation: $\vec r^{\prime}_{T}=\theta\vec r_{T}=\sqrt{1-\alpha^2}\vec r_{T}$, where $\alpha=V/v$. Such a non-classical effect occurs
only due to the existence of a minimum speed $V$ given for any direction in the space, so that even the transversal direction could not be
neglected, i.e., in SSR there should be a transformation for the vector at the transversal direction of the motion. Therefore, only if the
 speed $v$ is closer to $V$, a drastic dilation of $r_{T}$ occurs, that is to say
$r_{T}\rightarrow\infty$ when $v\rightarrow V$. Such a transversal dilation that occurs only close to the ultra-referential $S_V$ shows us
the 3-dimensional aspect of the background frame connected to $S_V$, which could not be simply reduced to 1-dimensional space since $V$
should remain invariant for any direction in the space.

 So, at the frame $S^{\prime}$, we have

\begin{equation}
\vec r^{\prime}=\vec r^{\prime}_{||}+\theta\vec r_{T},
\end{equation}
where $\vec r^{\prime}_{T}=\theta\vec r_{T}$.

As the direction of motion ($\vec r_{||}$) transforms in a similar way to the equation (1) for 1-dimensional case, we simply write (7),
as follows:

\begin{equation}
\vec r^{\prime}=\theta[\vec r_{T}+\gamma(\vec r_{||}-\vec v(1-\alpha)t],
\end{equation}
where we simply have $\vec r^{\prime}_{||}=\theta\gamma(\vec r_{||}-\vec v(1-\alpha)t)$, being $\vec r_{||}$ parallel to $\vec v$.
We have $\Psi=\theta\gamma$ and $\alpha=V/v$.

 From (8), we can see that $\theta$ appears as a multiplicative factor. Of course, if we make $\alpha=0$ ($V=0$) in (8), this implies
 $\theta=1$ and so we recover the well-known Lorentz tranformation of the 3-vector.

  We write $\vec r_T=\vec r-\vec r_{||}$. So by introducing this information into (8) and performing the calculations, we find:

\begin{equation}
\vec r^{\prime}=\theta[\vec r + (\gamma-1)\vec r_{||}-\gamma\vec v(1-\alpha)t]
\end{equation}

 We have $\vec r_{||}=rcos\phi\vec e_{||}$, where $\vec e_{||}$ is the unitary vector in the direction of motion, i.e.,
$\vec e_{||}=\frac{\vec v}{v}$. Angle $\phi$ is formed between the direction of $\vec r$ and $\vec e_{||}$. On the other hand, we can
 write: $rcos\phi=(\vec r.\vec v)/v$. Finally we get
$\vec r_{||}=\frac{(\vec r.\vec v)}{v^2}\vec v$. So we write (9) in the following way:

\begin{equation}
\vec r^{\prime}=\theta[\vec r + (\gamma-1)\frac{(\vec r.\vec v)}{v^2}\vec v-\gamma\vec v(1-\alpha)t]
\end{equation}

Transformation (10) above represents the 3-vector transformation in $(3+1)D$ space-time.

From (10), we can verify that, if we consider $\vec v$ to be in the same direction of $\vec r$, being $r\equiv X$, we obtain
$\frac{(\vec r.\vec v)}{v^2}\vec v=X\frac{\vec v}{v}=X\vec e_x$. So the transformation (10) is reduced to
 $x^{\prime}=\theta[X + (\gamma -1)X - \gamma v(1-\alpha)t]=\theta\gamma(X - v(1-\alpha)t)$, where $\Psi=\theta\gamma$. Such a
transformation is exactly the transformation (1) for the case of $(1+1)D$ space-time.

 Now we can realize that the generalization of the transformation (2) for the case of $(3+1)D$ space-time leads us to write:

 \begin{equation}
t^{\prime}=\theta\gamma[t - \frac{(\vec r.\vec v)}{c^2}(1-\alpha)],
\end{equation}
where $\theta\gamma=\Psi$. It is easy to verify that, if we have $\vec v||\vec r(\equiv X\vec e_x)$, we recover the time transformation (2)
for $(1+1)D$ space-time.

 Finally, by putting (11) and (10) in a matricial compact form, we find the following compact matrix:

\begin{equation}
\displaystyle\Omega_{4X4}=
\begin{pmatrix}
\theta\gamma & -\frac{\theta\gamma {\bf v}^T(1-\alpha)}{c} \\
-\frac{\theta\gamma{\bf v}(1-\alpha)}{c} & \left[\theta I+\theta(\gamma-1)\frac{{\bf v}{\bf v^T}}{v^2}\right]
\end{pmatrix},
\end{equation}
where $I=I_{3X3}$ is the identity matrix $(3X3)$ and ${\bf v}^T=(v_x, v_y, v_z)$ is the transpose of ${\bf v}$.

Now, in order to obtain the general inverse transformations, we should generalize the inverse transformations (4) and (5) for the case of $(3+1)D$ space-time.
To do that, we firstly consider the following known relations: $\vec r=\vec r_{||}+\vec r_T$ (12.a) and $\vec r^{\prime}=\vec r^{\prime}_{||}+\theta\vec r_T$
 (12.b), being $\theta\vec r_T=\vec r^{\prime}_T$ or $\vec r_T=\theta^{-1}\vec r^{\prime}_T$ (12.c).

In the direction of motion, the inverse transformation has the same form of (4), where we simply replace $X$ by $\vec r_{||}$ and
 $x^{\prime}$ by $\vec r^{\prime}_{||}$. So we write:

\begin{equation}
\vec r_{||}=\Psi^{\prime}[\vec r^{\prime}_{||}+\vec v(1-\alpha)t^{\prime}],
\end{equation}
where we have shown $\Psi^{\prime}=\Psi^{-1}/[1-\beta^2(1-\alpha)^2]\neq\Psi$.

Introducing (13) and (12.c) into (12.a), we find

\begin{equation}
\vec r=\theta^{-1}\vec r^{\prime}_T + \Psi^{\prime}[\vec r^{\prime}_{||}+\vec v(1-\alpha)t^{\prime}]
\end{equation}

As we have $\vec r^{\prime}_T=\vec r^{\prime}-\vec r^{\prime}_{||}$ (at the frame $S^{\prime}$), so introducing this relation into (14) and performing the calculations, we get

\begin{equation}
\vec r=\theta^{-1}\vec r^{\prime} + (\Psi^{\prime}-\theta^{-1})\vec r^{\prime}_{||}+\Psi^{\prime}\vec v(1-\alpha)t^{\prime},
\end{equation}
where we have $\vec r^{\prime}_{||}=(\frac{\vec r^{\prime}.\vec v}{v^2})\vec v$, and so we write

\begin{equation}
\vec r=\theta^{-1}\vec r^{\prime} + (\Psi^{\prime}-\theta^{-1})(\frac{\vec r^{\prime}.\vec v}{v^2})\vec v +\Psi^{\prime}\vec v(1-\alpha)t^{\prime},
\end{equation}

As we already know $\Psi^{\prime}=\Psi^{-1}/[1-\beta^2(1-\alpha)^2]=\theta^{-1}\gamma^{-1}/[1-\beta^2(1-\alpha)^2]$, we can also write (16) in the following way:

\begin{equation}
\vec r=\theta^{-1}\vec r^{\prime} + \theta^{-1}\left[\left(\frac{\gamma^{-1}}{1-\beta^2_*}-1\right)(\frac{\vec r^{\prime}.\vec v}{v^2})+
\frac{(\gamma^{-1})_*}{1-\beta^2_*}t^{\prime}\right]\vec v,
\end{equation}
where we have used the simplified notation $\beta_*=\beta(1-\alpha)$. We also have $(\gamma^{-1})_*=\gamma^{-1}(1-\alpha)$.

 Now it is natural to conclude that the time inverse transformation is given as follows:

\begin{equation}
t=\frac{\theta^{-1}\gamma^{-1}}{1-\beta^2(1-\alpha)^2}\left[t^{\prime}+ \frac{\vec r^{\prime}.\vec v}{c^2}(1-\alpha)\right]
\end{equation}

In (17) and (18), if we make $\alpha=0$ (or $V=0$), we recover the $(3+1)D$ Lorentz inverse transformations.

From (18) and (17) we get the following compact inverse matrix of transformation:

\begin{equation}
\displaystyle\Omega_{4X4}^{-1}=
\begin{pmatrix}
\frac{\theta^{-1}\gamma^{-1}}{1-\beta^2_*} & \frac{\theta^{-1}\gamma^{-1}{\bf v}^T_*}{c(1-\beta^2_*)} \\
\frac{\theta^{-1}\gamma^{-1}{\bf v_*}}{c(1-\beta^2_*)} &
\left[\theta^{-1}I+\theta(\frac{\gamma^{-1}}{1-\beta^2_*}-1)\frac{{\bf v}{\bf v^T}}{v^2}\right]
\end{pmatrix},
\end{equation}
where ${\bf v}^T_*={\bf v}^T(1-\alpha)$, ${\bf v}_*={\bf v}(1-\alpha)$ and $\beta_*=\beta(1-\alpha)$.

 Now we can compare the inverse matrix (19) with (12) and also verify that $\Omega^{-1}_{4X4}\neq\Omega^T_{4X4}$, in a similar way as made
before for the case $(1+1)D$.

\section{\label{sec:level1} Flat space-time with the ultra-referential $S_V$}

\subsection{Flat space-time in SR}

First of all, as it is well-known, according to SR, the space-time interval is

\begin{equation}
ds^2=g_{\mu\nu}dx^{\mu}dx^{\nu}=c^2dt^2-dx^2-dy^2-dz^2,
\end{equation}
where $g_{\mu\nu}$ is the Minkowski metric of the flat space-time.

 Due to the invariance of the norm of the 4-vector, we have $ds^2$ (frame $S$) $=ds^{\prime 2}$ (frame $S^{\prime}$).
 By considering a moving particle with a speed $v$, being on the origin of $S^{\prime}$, we write

\begin{equation}
ds^2=c^2dt^2-dx^2-dy^2-dz^2\equiv ds^{\prime 2}=c^2d\tau^2,
\end{equation}
from where we extract the following relation between time intervals:

\begin{equation}
\Delta\tau=\Delta t\left[1-\frac{(dx^2+dy^2+dz^2)}{c^2dt^2}\right]^{\frac{1}{2}}=\Delta t\sqrt{1-\frac{v^2}{c^2}}
\end{equation}

Fixing the proper time interval $\Delta\tau$, thus for $v\rightarrow c$, this leads to the drastic increasing of the improper time interval
 ($\Delta t\rightarrow\infty$). This is the well-known {\it time dilatation}.

\subsection{Flat space-time in SSR}

Due to the non-locality of the ultra-referential $S_V$ connected to a background field that fills uniformly the whole flat space-time,
 when the speed $v$ ($S^{\prime}$) of a particle is much closer to $V$ ($S_V$), a very drastic dilatation of the proper space-time interval $dS^{\prime}$
 occurs. In order to describe such an effect in terms of metric, let us write:

\begin{equation}
dS^{\prime 2}_v=dS_v^2=\Theta_v ds^2=\Theta_v g_{\mu\nu}dx^{\mu}dx^{\nu},
\end{equation}
where $dS_v^{\prime}~(=dS^{\prime})$ is the dilated proper space-time interval (in $S^{\prime}$) due to the dilatation factor (function) $\Theta_v$, which
 depends on the speed $v$, so that $\Theta_v$ diverges ($\rightarrow\infty$) when $v\rightarrow V$, and thus $\Delta S^{\prime}_v=\Delta S_v>>\Delta s$
 ($\Delta S_v\rightarrow\infty$), breaking strongly the invariance
 of $\Delta s$. On the other hand, when $v>>V$ we recover $\Delta s$, i.e., $\Delta S^{\prime}_v=\Delta S_v\approx\Delta s$, which does not depend on $v$
 since $\Theta_v\approx 1$ (approximation for SR theory). So considering such conditions, let us write

\begin{equation}
\Theta_v=\Theta (v)=\frac{1}{(1-\frac{V^2}{v^2})},~[3]
\end{equation}
which leads to an effective (deformed) metric $G_{(v)\mu\nu}=\Theta(v)g_{\mu\nu}$ due to the dilatation factor $\Theta_v$. So we have
$dS_v^2=G_{(v)\mu\nu}dx^{\mu}dx^{\nu}$. We observe that $\Theta(v)=\theta(v)^{-2}$, where we have shown that $\theta(v)=\sqrt{1-\frac {V^2}{v^2}}$ (section 2).
 Actually the dilatation factor $\Theta_v$ appears due to the presence of the privileged frame $S_V$ as a background field being inherent to the deformed metric
 $G_{(v)\mu\nu}$. Thus the transformations in such a space-time of SSR do not necessarily form a group. This subject will be treated
in a further work.

 The presence of the dilatation factor $\Theta_v$ affects directly the proper time of the moving particle at $S^{\prime}$, which becomes a variable
parameter in SSR, in the sense that, just close to $V$, there emerges a dilatation of the proper time interval $\Delta\tau$ in relation
to the improper one $\Delta t$, namely $\Delta\tau>\Delta t$. In short, such a new relativistic effect in SSR shows us that the proper time interval becomes
a variable and deformable parameter connected to the motion $v$, as well as the improper time interval is deformable, namely
the so-called {\it time dilatation}.

  In SSR, due to the connection between the proper time interval and the motion, let us call $\Delta\tau_v$ (at $S^{\prime}$) to represent
 an intrinsic variable
 of proper time interval depending on the motion $v$. Of course for $v>>V$, we expect that such a dependence can be neglected,
recovering the proper time of SR. But, close to $V$, the new effect of the dilatation of $\Delta\tau_v$ in relation to $\Delta t$ ($\Delta\tau_v>\Delta t$)
emerges, and it is due to the dilatation factor $\Theta(v)$. So according to (23) we find the following equivalence of dilated space-time intervals:

\begin{equation}
dS_v^2=\Theta_v[c^2dt^2-dx^2-dy^2-dz^2]\equiv dS_v^{\prime 2}=c^2d\tau_v^2,
\end{equation}
 being $\Theta_v=\left(1-\frac{V^2}{v^2}\right)^{-1}$. Here we have made $dx^{\prime 2}=dy^{\prime 2}=dz^{\prime 2}=0$. If we make $V\rightarrow 0$ (no
 ultra-referential $S_V$) or even $v>>V$ ($\Theta_v\approx 1$), we recover the well-known equivalence (invariance) of intervals in SR (see
 (21)).

 As the deformed metric $G_{\mu\nu}(v)$ of SSR depends on velocity, it seems to be related to a kind of Finsler metric, namely a Finslerian
(non-Riemannian) space with a metric depending on position and velocity, that is, $g_{\mu\nu}(x,\dot{x})$\cite{23}\cite{24}
 \cite{25}\cite{26}\cite{27}. Of course, if there is no dependence on velocity, the Finsler space turns out to be a Riemannian space. Such
 a possible connection between $G_{\mu\nu}(v)$ and Finslerian geometry should be investigated.

 From (25) we obtain

\begin{equation}
d\tau^2_v\left(1-\frac{V^2}{v^2}\right)=dt^2\left(1-\frac{v^2}{c^2}\right),
\end{equation}
which finally leads to

\begin{equation}
\Delta\tau\sqrt{1-\frac{V^2}{v^2}}=\Delta t\sqrt{1-\frac{v^2}{c^2}}
\end{equation}

 Equation 27 reveals a perfect symmetry ($V<v<c$) in the sense that both intervals of time $\Delta t$ and $\Delta\tau$ can dilate, namely
 $\Delta t$ dilates for $v\rightarrow c$ and, on the other hand, $\Delta\tau$ dilates for $v\rightarrow V$. But, if $V\rightarrow 0$, we
 break such a symmetry of SSR and so we recover the well-known time equation (eq.22) of SR, where only $\Delta t$ dilates.

 For the sake of simplicity, we simply use the notation $\Delta\tau$ ($=\Delta\tau_v$) for representing the proper time interval in the
 time equation of SSR (eq.27).

From (27) we notice that, if we make $v=v_0=\sqrt{cV}$ (a geometric average between $c$ and $V$), we exactly find the equality $\Delta\tau~(S^{\prime})=
\Delta t~(S)$, namely a newtonian result where the time intervals are the same. Thus we conclude that $v_0$ represents a special intermediate speed in SSR
 ($V<<v_0<<c$) such that, if:

a) $v>>v_0$ (or even $v\rightarrow c$), we get $\Delta\tau<<\Delta t$. This is the well-known {\it improper time dilatation}.

b) $v<<v_0$ (or even $v\rightarrow V$), we get $\Delta\tau>>\Delta t$. Let us call such a new effect as {\it improper time contraction} or
{\it dilatation of the proper time interval $\Delta\tau$ in relation to the improper time interval $\Delta t$}. This effect is
more evident only for $v\rightarrow V$, so that we have $\Delta\tau\rightarrow\infty$ for $\Delta t$ fixed (see eq.27). In other words this means that the proper
time elapses faster than the improper one closer to $V$.

 In SSR, it is interesting to notice that we restore the newtonian regime when $V<<v<<c$, which represents an intermediary regime of speeds
 so that we get the newtonian approximation from equation 27, i.e., $\Delta\tau\approx\Delta t$.

 Equation 27 can be written in the form

\begin{equation}
c^2\Delta\tau^2=\frac{1}{(1-\frac{V^2}{v^2})}[c^2\Delta t^2-v^2\Delta t^2]
\end{equation}

By placing eq.28 in a differential form and manipulating it, we will obtain

\begin{equation}
c^2\left(1-\frac{V^2}{v^2}\right)\frac{d\tau^2}{dt^2} + v^2 = c^2
\end{equation}

 Equation (29) shows us that both of the speeds related to the marching of time  (``temporal-speed''
 $v_t=c\sqrt{1-\frac{V^2}{v^2}}\frac{d\tau}{dt}$)
 and the spatial speed $v$ form the vertical and horizontal legs of a rectangular triangle respectively (Fig.2). The hypotenuse of the
 triangle is $c=(v_t^2+v^2)^{1/2}$ representing the spatio-temporal speed of any particle.

\begin{figure}
\includegraphics[scale=0.9]{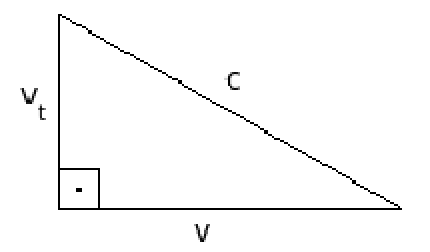}
\caption{We see that the horizontal leg represents the spatial-speed $v$ , while the vertical leg represents the temporal-speed $v_t$
 (march of time),
 where $v_t=\sqrt{c^2-v^2}=c\sqrt{1-v^2/c^2}=c\sqrt{1-V^2/v^2}d\tau/dt$ (see eq.27), so that we always have $v^2+v_t^2=c^2$. In SR, when
 $v=0$,
  the horizontal leg vanishes (no spatial speed) and so the vertical leg becomes maximum ($v_t=v_{tmax}=c$). However, according to SSR, due
 to the existence
of a minimum limit of spatial speed ($V$), we can never nullify the horizontal leg, so that the maximum temporal speed (maximum vertical leg) is
$v_{tmax}=\sqrt{c^2-V^2}=c\sqrt{1-V^2/c^2}<c$. On the other hand $v_t$ (the vertical leg) cannot be zero since $v=c$ is forbidden for massive particles.
 So we conclude that a rectangular triangle is always preserved since both temporal and spatial speeds cannot vanish and so they always
 coexist, providing a strong symmetry of SSR.}
\end{figure}

 Looking at Fig.2 we should consider three important cases, namely:

 a)    If $v\approx c$, $v_t\approx 0$ (the marching of the time is very slow), so that $\Psi>>1$, leading to $\Delta t>>\Delta\tau$ ({\it
 dilatation of the improper time}).

 b)    If $v=v_0=\sqrt{cV}$, $v_t=\sqrt{c^2-v_0^2}$ (the marching of the time is fast), so that $\Psi=\Psi_0=\Psi(v_0)=1$, leading to
 $\Delta\tau=\Delta t$.

 c)    If $v\approx V(<<v_0)$, $v_t\approx\sqrt{c^2-V^2}=c\sqrt{1-V^2/c^2}$ (the marching of the time is even faster than that one at $S$),
 so that $\Psi<<1$, leading to $\Delta t<<\Delta\tau$ ({\it contraction of the improper time or dilatation of the proper time with respect
 to the improper one}).

\section{\label{sec:level1} Relativistic dynamics in SSR}

\subsection{Energy and momentum}

Let us firstly define the 4-velocity in the presence of $S_V$, as follows:

\begin{equation}
 U^{\mu}=\left[\frac{\sqrt{1-\frac{V^2}{v^2}}}{\sqrt{1-\frac{v^2}{c^2}}}~ , ~
\frac{v_{\alpha}\sqrt{1-\frac{V^2}{v^2}}}{c\sqrt{1-\frac{v^2}{c^2}}}\right],
\end{equation}
where $\mu=0,1,2,3$ and $\alpha=1,2,3$. If $V\rightarrow 0$, we recover the well-known 4-velocity of SR. From (30) it is interesting to observe that
the 4-velocity of SSR vanishes in the limit of $v\rightarrow V$ ($S_V$), i.e., $U^{\mu}=(0,0,0,0)$, whereas in SR, for $v=0$ we find $U^{\mu}=(1,0,0,0)$.

 The 4-momentum is
\begin{equation}
 p^{\mu}=m_0cU^{\mu},
   \end{equation}
being $U^{\mu}$ given in (30). So we find

\begin{equation}
 p^{\mu}=\left[\frac{m_0c\sqrt{1-\frac{V^2}{v^2}}}{\sqrt{1-\frac{v^2}{c^2}}}~ , ~
\frac{m_0v_{\alpha}\sqrt{1-\frac{V^2}{v^2}}}{\sqrt{1-\frac{v^2}{c^2}}}\right],
\end{equation}
where $p^0=E/c$, such that

\begin{equation}
E=cp^0=mc^2=m_0c^2\frac{\sqrt{1-\frac{V^2}{v^2}}}{\sqrt{1-\frac{v^2}{c^2}}},
\end{equation}
where $E$ is the total energy of the particle with speed $v$ in relation to the absolute inertial frame (ultra-referential $S_V$). From
 (33), we observe that, if
$v\rightarrow c\Rightarrow E\rightarrow\infty$. If $v\rightarrow V\Rightarrow E\rightarrow 0$ and if $v=v_0=\sqrt{cV}\Rightarrow E=E_0=m_0c^2$ (proper energy
in SSR). Figure 3 shows us the graph for the energy $E$ in eq.33.

\begin{figure}
\begin{center}
\includegraphics[scale=0.7]{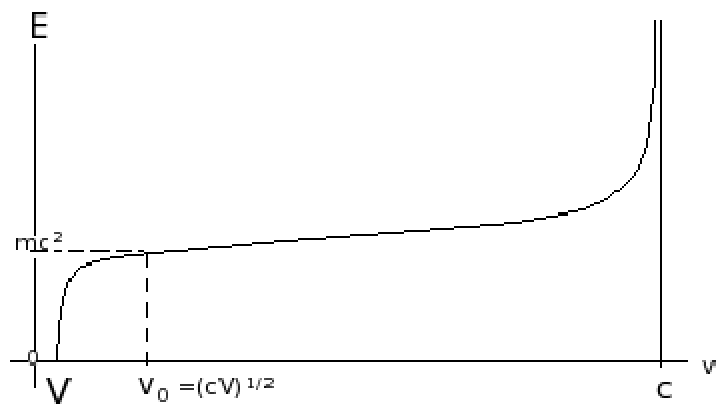}
\end{center}
\caption{$v_0=\sqrt{cV}$ is a speed such that we get the proper energy of the particle ($E_0=m_0c^2$) in SSR,
  where $\Psi_0=\Psi(v_0)=1$. For $v<<v_0$ or closer to $S_V$ ($v\rightarrow V$), a new relativistic correction on energy arises, so that
 $E\rightarrow 0$. On the other hand, for $v>>v_0$, being $v\rightarrow c$, so we find $E\rightarrow\infty$.}
\end{figure}

From (32) we also obtain the (spatial) momentum, namely:

\begin{equation}
\vec p = m_0\vec v\frac{\sqrt{1-\frac{V^2}{v^2}}}{\sqrt{1-\frac{v^2}{c^2}}},
\end{equation}
where $\vec v=(v_1,v_2,v_3)$.

From (32), performing the quantity $p^{\mu}p_{\mu}$, we obtain the energy-momentum relation of SSR, as follows:

\begin{equation}
p^{\mu}p_{\mu}=\frac{E^2}{c^2}-\vec p^2=m_0^2c^2\left(1-\frac{V^2}{v^2}\right),
\end{equation}
where $\vec p^2=p_1^2+p_2^2+p_3^2$. 

From (35) we find

\begin{equation}
E^2=c^2p^2+m_0^2c^4\left(1-\frac{V^2}{v^2}\right)
\end{equation}

In the present work, as we are focusing our attention on some dynamical implications of a minimum speed, let us leave a more detailed
 development of the physical consequences of SSR in terms of field-theory actions and gravitational extensions to be explored elsewhere.

\subsection{Power of an applied force: the energy barrier of the minimum speed $V$}

Let us consider a force applied to a particle, in the same direction of its motion. More general cases where the force is not necessarily parallel
to velocity will be treated elsewhere. In our specific case ($\vec F||\vec v$), the relativistic power $P_{ow}(=vdp/dt)$ is given as follows:

\begin{equation}
P_{ow}=v\frac{d}{dt}\left[m_0v\left(1-\frac{V^2}{v^2}\right)^{\frac{1}{2}}\left(1-\frac{v^2}{c^2}\right)^{-\frac{1}{2}}\right],
\end{equation}
where we have used the momentum $p$ given in (34).

After performing the calculations in (37), we find

\begin{equation}
 P_{ow}=\left[\frac{\left(1-\frac{V^2}{v^2}\right)^{\frac{1}{2}}}{\left(1-\frac{v^2}{c^2}\right)^{\frac{3}{2}}}
+\frac{V^2}{v^2\left(1-\frac{v^2}{c^2}\right)^{\frac{1}{2}}\left(1-\frac{V^2}{v^2}\right)^{\frac{1}{2}}}\right]
\frac{dE_k}{dt},
\end{equation}
where $E_k=\frac{1}{2}m_0v^2$.

If we make $V\rightarrow 0$ and $c\rightarrow\infty$ in (38), we simply recover the power obtained in newtonian mechanics, namely
$P_{ow}=dE_k/dt$. Now, if we just consider $V\rightarrow 0$ in (38), we recover the well-known relativistic power (SR), namely
$P_{ow}=(1-v^2/c^2)^{-3/2}dE_k/dt$. We notice that such a relativistic power tends to infinite ($P_{ow}\rightarrow\infty$) in the
limit $v\rightarrow c$. We explain this result as an effect of the drastic increase of an effective inertial mass close to $c$, namely
$m_{eff}=m_0(1-v^2/c^2)^{k^{\prime\prime}}$, where $k^{\prime\prime}=-3/2$. We must stress that such an effective inertial mass is the response to an applied
force parallel to the motion according to Newton second law, and it increases faster than the relativistic mass $m=m_r=m_0(1-v^2/c^2)^{-1/2}$.

  The effective inertial mass $m_ {eff}$ we have obtained is a longitudinal mass $m_L$, i.e., it is a response to the force applied in the
 direction of motion.
  In SR, for the case where the force is perpendicular to velocity, we can show that the transversal mass increases like the relativistic
 mass, i.e.,
 $m=m_T=m_0(1-v^2/c^2)^{-1/2}$, which differs from the longitudinal mass $m_L=m_0(1-v^2/c^2)^{-3/2}$.  So in this sense there is anisotropy
 of the effective inertial mass to be also investigated in more details by SSR in a further work.

 The mysterious discrepancy between the relativistic mass $m$ ($m_r$) and the longitudinal inertial mass $m_L$ from Newton second law
 (eq.38) is a
 controversial issue\cite{28}\cite{29}\cite{30}\cite{31}\cite{32}\cite{33}. Actually the newtonian notion about inertia as the resistance
 to acceleration
($m_L$) is not compatible with the relativistic dynamics ($m_r$) in the sense that we generally cannot consider $\vec F=m_{r}\vec a$. The dynamics of SSR
 aims to give us a new interpretation for the inertia of the newtonian point of view in order to make it compatible with the relativistic
 mass. This compatibility
 will be possible just due to the influence of the background field that couples to the particle and ``dresses" its relativistic mass in
 order to generate an
 effective (dressed) mass in accordance with the newtonian notion about inertia (from eqs.37 and 38). This issue will be clarified in this
 section.

From (38), it is important to observe that, when we are closer to $V$, there emerges a completely new result (correction) for power, namely:

\begin{equation}
P_{ow}\approx\left(1-\frac{V^2}{v^2}\right)^{-\frac{1}{2}}\frac{d}{dt}\left(\frac{1}{2}m_0v^2\right),
\end{equation}
given in the approximation $v\approx V$. So we notice that $P_{ow}\rightarrow\infty$ when $v\approx V$. We can also make the limit
$v\rightarrow V$ for the general case (eq.38) and so we obtain an infinite power ($P_{ow}\rightarrow\infty$). Such a new relativistic effect deserves
the following very important comment:  Although we are in the limit of very low energies close to $V$, where the energy of the particle ($mc^2$) tends
to zero according to the approximation $E=mc^2\approx m_0c^2(1-V^2/v^2)^{k}$ with $k=1/2$ (make the approximation $v\approx V$ for eq.33),
 on the
other hand the power given in (39) shows us that there is an effective inertial mass that increases to infinite in the limit $v\rightarrow V$, that is to
say, from (39) we get the effective mass $m_{eff}\approx m_0(1-V^2/v^2)^{k^{\prime}}$, where $k^{\prime}=-1/2$. Therefore, from a dynamical point of view,
 the negative exponent $k^{\prime}$ ($=-1/2$) for power at very low velocities (eq.39) is responsible for the inferior barrier of the
 minimum speed $V$, as well as the exponent $k^{\prime\prime}=-3/2$ of the well-known relativistic power is responsible for the top barrier
 of the speed of light $c$ according to Newton second law.

 In order to see clearly both exponents $k^{\prime}=-1/2$ (inferior inertial barrier $V$) and $k^{\prime\prime}=-3/2$ (top inertial barrier
 $c$), let us write the general formula of power (eq.38) in the following alternative way after some algebraic manipulations on it,
  namely:

\begin{equation}
P_{ow}=\left(1-\frac{V^2}{v^2}\right)^{k^{\prime}}\left(1-\frac{v^2}{c^2}\right)^{k^{\prime\prime}}\left(1-\frac{V^2}{c^2}\right)
\frac{dE_k}{dt},
\end{equation}
where $k^{\prime}=-1/2$ and $k^{\prime\prime}=-3/2$. Now it is easy to see that, if $v\approx V$ or even $v<<c$, eq.40 recovers the
 approximation (39). As $V<<c$, the ratio $V^2/c^2$ in (40) is a very small dimensionless constant\cite{2}. So it could be neglected.

From (40) we get the effective inertial mass $m_{eff}$ of SSR, namely:

\begin{equation}
m_{eff}=m_0\left(1-\frac{V^2}{v^2}\right)^{-\frac{1}{2}}\left(1-\frac{v^2}{c^2}\right)^{-\frac{3}{2}}\left(1-\frac{V^2}{c^2}\right)
\end{equation}

We must stress that $m_{eff}$ in (41) is a longitudinal mass $m_L$. The problem of mass anisotropy will be treated elsewhere, where we
will intend to show that, just for the approximation $v\approx V$, the effective inertial mass becomes practically isotropic, that is to
 say
$m_L\approx m_T\approx m_0\left(1-\frac{V^2}{v^2}\right)^{-1/2}$. This important result will show us the isotropic aspect of the
 vacuum-$S_V$
so that the inferior barrier $V$ has the same behavior of response ($k^{\prime}=-1/2$) for any direction in the space, namely for any angle
 between the applied force and the velocity of the particle.

 We must point out the fact that $m_{eff}$ has nothing to do with the ``relativistic mass" (relativistic energy $E$ in eq.33) in the sense
 that
$m_{eff}$ is dynamically responsible for both barriers $V$ and $c$. The discrepancy between the ``relativistic mass" (energy $mc^2$ of the particle) and
such an effective inertial mass ($m_{eff}$) can be interpreted under SSR theory, as follows: $m_{eff}$ is a dressed inertial mass given in response to the presence
of the vacuum-$S_V$ that works like a kind of ``fluid" in which the particle $m_0$ is immersed, while the ``relativistic mass" in SSR 
 (eq.33) works like a
 bare inertial mass in the sense that it is not considered to be under the dynamical influence of the ``fluid" connected to the
 vacuum-$S_V$. That is the reason
 why the exponent $k=1/2$ in eq.33 cannot be used to explain the existence of an infinite inferior barrier at $V$, namely the vacuum-$S_V$
 barrier is governed by
the exponent $k^{\prime}=-1/2$ as shown in (39), (40) and (41), which prevents that $v_*(=v-V)\leq 0$.

 The difference betweeen the dressed (effective) mass and the relativistic (bare) mass, i.e., $m_{eff}-m$ represents an interactive
 increment of mass $\Delta m_{i}$ that has purely origin from the vacuum energy-$S_V$, mamely

\begin{equation}
\Delta m_{i}= m_0\left[\frac{\left(1-\frac{V^2}{c^2}\right)}{\left(1-\frac{V^2}{v^2}\right)^{\frac{1}{2}}
\left(1-\frac{v^2}{c^2}\right)^{\frac{3}{2}}}- \frac{\left(1-\frac{V^2}{v^2}\right)^{\frac{1}{2}}}{\left(1-\frac{v^2}{c^2}\right)^{\frac{1}{2}}}\right]
\end{equation}

We have $\Delta m_{i}=m_{eff}-m$, being $m_{eff}=m_{dressed}$ given in eq.41 and $m$ ($m_r$) given in eq.33.

 From (42) we consider the following special cases:

a) for $v\approx c$ we have

\begin{equation}
 \Delta m_{i}\approx m_0\left[\left(1-\frac{v^2}{c^2}\right)^{-\frac{3}{2}}-\left(1-\frac{v^2}{c^2}\right)^{-\frac{1}{2}}\right]
\end{equation}

As the effective inertial mass $m_{eff}$ ($m_L$) increases much faster than the bare (relativistic) mass $m$ ($m_r$) close to the speed $c$,
 there is an increment of inertial mass $\Delta m_i$ that dresses $m$ in direction of its motion and it tends to be infinite when $v\rightarrow c$,
i.e., $\Delta m_i\rightarrow\infty$.

b) for $V<<v<<c$ (newtonian or intermediary regime) we find $\Delta m_i\approx 0$, where we simply have $m_{eff}$ ($m_{dressed}$)$\approx m\approx m_0$.
 This is the classical case.

c) for $v\approx V$ (close to the vacuum-$S_V$ regime) we have the following approximation:

\begin{equation}
\Delta m_{i}=(m_{dressed}-m)\approx m_{dressed}\approx\frac{m_0}{\sqrt{1-\frac{V^2}{v^2}}},
\end{equation}
where $m\approx 0$ when $v\approx V$ (see eq.33).

It is interesting to compare (44) ($m_{dressed}\approx\theta^{-1}m_0$) with the transformation given for the transversal direction, namely
$r_T=\theta^{-1}r_T^{\prime}$, so that we find $\frac{m_{dressed}}{m_0}\approx\frac{r_T}{r_T^{\prime}}=\theta^{-1}$, which implies
$m_0r_T\approx m_{dressed}r_T^{\prime}$. If $v\rightarrow V$, this leads to
$m_{dressed}\rightarrow\infty\Rightarrow r_T\rightarrow\infty$ for $r^{\prime}_T$ fixed to be finite and $m_0>0$, so that we note that
 $m_{dressed}$ and $r_T$ are directly related to each other.

 The approximation (44) shows that the whole dressed mass has purely origin from the energy of vacuum-$S_V$, being $m_{dressed}$ the pure
 increment
$\Delta m_{i}$, since the bare (relativistic) mass $m$ of the own particle
 almost vanishes in such a regime ($v\approx V$), and thus an inertial effect only due to the vacuum (``fluid")-$S_V$
remains. We see that $\Delta m_{i}\rightarrow\infty$ when $v\rightarrow V$. In other words, we can interpret this infinite barrier of vacuum-$S_V$ by
considering the particle to be strongly coupled to the background field-$S_V$ for all directions of the space . The isotropy of $m_{eff}$ in this regime will
be shown in detail elsewhere, being $m_{eff}=m_L=m_T\approx m_0(1-V^2/v^2)^{-1/2}$. In such a regime the particle practically loses its locality
(``identity") in the sense that it is spread out isotropically in the whole space and it becomes strongly coupled to the vacuum field-$S_V$, leading to an
 infinite value of $\Delta m_{i}$. Such a divergence has origin from the dilatation factor $\Theta_v(\rightarrow\infty)$ for this regime
 ($v\approx V$), so that we can rewrite (44) in the following way: $\Delta m_{i}\approx m_{dressed}\approx m_0\Theta(v)^{1/2}$.

 Figure 4 shows the graph for the longitudinal effective inertial mass $m_{eff}=m_L$ ($m_{dressed}$) as a function of the speed $v$,
 according to equation 41.

\begin{figure}
\begin{center}
\includegraphics[scale=0.75]{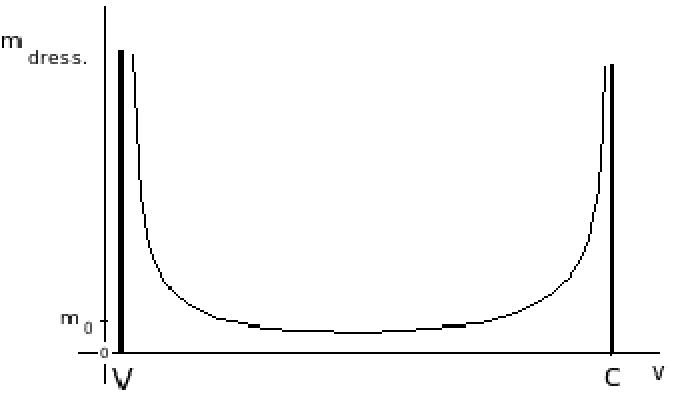}
\end{center}
\caption{The graph shows us two infinite barriers at $V$ and $c$, providing an aspect of symmetry of SSR. The first barrier ($V$) is exclusively due
 to the vacuum-$S_V$, being interpreted as a barrier of pure vacuum energy. In this regime we have the following approximations:
 $m_{eff}=m_{dressed}\approx\Delta m_{i}\approx m_0(1-V^2/v^2)^{-1/2}$ and  $m_r\approx m_0(1-V^2/v^2)^{1/2}$ (see Fig.3), so that
 $m_{dressed}\rightarrow\infty$
 and $m=m_r=m_{bare}\rightarrow 0$ when $v\rightarrow V$. The second barrier ($c$) is a sum (mixture) of two contributions, namely the own
 bare
 (relativistic) mass $m$ that increases with the factor $\gamma=(1-v^2/c^2)^{-1/2}$ (see Fig 3) plus the interactive increment $\Delta
 m_{i}$ due to the
 vacuum energy-$S_V$, so that $m_{dressed}=m_L=m+\Delta m_{i}\approx m_0(1-v^2/c^2)^{-3/2}$. This is a longitudinal effect. For the
 transversal effect,
  $\Delta  m_{i}=0$ since we get $m_T=m$. This result will be shown elsewhere.}
\end{figure}

 Let us now consider the de-Broglie wavelength of a particle, namely:

\begin{equation}
\lambda=\frac{h}{P}=\frac{h}{m_0v}\frac{\sqrt{1-\frac{v^2}{c^2}}}{\sqrt{1-\frac{V^2}{v^2}}},
 \end{equation}
from where we have used the momentum given in eq.21.

If $v\rightarrow c\Rightarrow\lambda\rightarrow 0$ ({\it spatial contraction}), and if $v\rightarrow V\Rightarrow\lambda\rightarrow\infty$
({\it spatial dilatation to the infinite}). This limit leads to an infinite dilatation factor, i.e., $\Theta_v\rightarrow\infty$ (see
 (24)),
 where the wavelength of the particle tends to infinite (see eq.45). So alternatively we can write eq.45 in the following way:
  $\lambda=\Theta_v^{1/2}(h/\gamma m_0v)$, where $h/\gamma m_0v$ represents the well-known de-Broglie wavelength with the relativistic
 correction
  for momentum, i.e., with the Lorentz factor $\gamma$. $\Theta_v$ is the dilatation factor that leads to a drastic dilatation of the
 wavelength close to $V$.

\section{\label{sec:level1} Foundations of the third law of thermodynamics according to the new relativistic dynamics}

\subsection{The classical model for an ideal gas}

 Consider a non-relativistic particle with mass $m_0$ and speed $v$ inside a cubic box with side $L$. As it is well-known, its motion $v$ generates
a ``pressure" $P$ on the internal walls of the box, namely:

\begin{equation}
P=\frac{pv}{L^3}=\frac{m_0v^2}{L^3},
\end{equation}
where $V_{ol}=L^3$ is the $3D$ volume of the box. $p=m_0v$ is the non-relativistic momentum.

 If we have a very large number $N$ of identical ``particles" (atoms or molecules) of an ideal gas with a temperature $T$ inside such a box, we write

\begin{equation}
PV_{ol}=N m_0\left<v^2\right>=\nu Nk_B T=\nu nRT,
\end{equation}
where $R=(N/n)k_B=N_ak_B$, being $n$ the number of moles, $N_a$ the Avogrado number, $k_B$ the Boltzmann constant and $R$ the universal constant of gases.
  We have the statistical average $\left<v^2\right>=\Sigma_{i=1}^N v_i^2/N$. $P$ is the pressure of the gas. $\nu$ represents the number of degrees of freedom
for each ``particle" (atom or molecule) inside the box. If the ``particle" is considered to be punctual (without intrinsic degrees of freedom), we have
 $\nu\equiv D$, which corresponds to the dimensionality of the system. In our case we have $\nu=D=3$ and so the mean energy per particle is
 $\left<\epsilon\right>=(1/2)m_0\left<v^2\right>=3(1/2)k_BT$.

 From (47) it is easy to see that, if we make $\left<v^2\right>=0$, this leads to $T=0$. So based on a purely dynamical aspect such as the
 classical
mechanics and even the relativistic mechanics, it would be really possible to admit the existence of the absolute zero temperature ($T=0K$). However,
according to
the thermodynamics viewpoint, the third law (a phenomenological law) prevents to attain $T=0K$. In this sense, the dynamical laws are not compatible
with thermodynamics although quantum mechanics postulates a zero point of energy due to the uncertainty principle in order to forbid a particle
to be at rest inside a box. Actually, in spite of the quantum principles, we aim to search for a purely dynamical and fundamental explanation for the third law
of thermodynamics. So let us consider the deformed relativistic dynamics with a minimum speed to deal with an ideal gas, being consistent
with quantum principles\cite{2}.

\subsection{The new relativistic model for an ideal gas}

Now consider a particle with relativistic momentum $p$ as given in eq.(34) ($p=m_0\Psi v$). As we are just interested in the new
 corrections for very low energies, we make the approximation in eq.(34) for $v<<c$ ($\Psi\approx\theta$), namely $p\approx m_0\theta v$.
 In this case, the momentum $p$ is connected to the ``pressure" $P$, as follows:

\begin{equation}
P=\frac{pv}{V_{ol}}\approx\frac{m_0\theta v^2}{V_{ol}},
\end{equation}
where $\theta =\theta(v)=\sqrt{1-\frac{V^2}{v^2}}$.

If we consider a very large number $N$ of identical particles of an ideal gas at low temperature, we write

\begin{equation}
PV_{ol}\approx N m_0\sqrt{1-\frac{V^2}{\left<v^2\right>}}\left<v^2\right>=3fNk_BT,
\end{equation}
where $f$ is a function of the temperature ($f(T)$) to be investigated. In the classical case ($T>>0K$) we get $\nu=D=3$, i.e., in a more
 particular case, where $V<<\sqrt{\left<v^2\right>}<<c$, we recover the
classical (newtonian) case as given in (47).

 Since we take into account the new relativistic effects only for very low velocities
close to $V$, we use the approximation (49), however we must warn that the energy equi-partition theorem does not work in this regime of condensation
at very low temperatures. Such a subject will be deeply explored in a further work, where we intend to show that the classical molar specific heat ($(3/2)R$)
is corrected with a function of temperature, namely $(3/2)f(T)R$, where $f(T)$ in (49) will be obtained, being $0<f(T)<1$. Due to a breakdown of the
energy equi-partition, $f(T)$ plays the role of making an effective reduction of the degrees of freedom, that is to say $\nu_{eff}=f(T)\nu$ ($\nu_{eff}<\nu$).
 So, in our case given in (49), we have an effective dimension $D_{eff}=3f(T)$. We expect $f(T)\approx 1$ for higher temperatures,
 recovering the classical case. Even
 so, since the function $f(T)$ will not affect the present analysis, let us simply write the following proportionality:

\begin{equation}
PV_{ol}\approx N m_0 \left<v^2\right>\sqrt{1-\frac{V^2}{\left<v^2\right>}}\propto Nk_B T
\end{equation}

 According to (50), if $T\rightarrow 0$K, thus $\sqrt{\left<v^2\right>}\rightarrow V$ and $P\rightarrow 0$. However, since we have already
 shown that the
minimum speed $V$ forms an unattainable and inferior barrier, we are able to explain from a dynamical viewpoint why the absolute zero
temperature becomes unattainable, that is to say $T$ tends to absolute zero ($T\rightarrow 0$K), but $T$ never attains the absolute zero
 ($0$K) due to the inferior energy barrier of the unattainable minimum speed $V$. In other words, we say that, as there is no zero speed in
 SSR, the absolute zero temperature ($T=0$K) must be directly related to a minimum (non zero) speed $V$, and since $V$ is an unattainable
limit, this can explain from a dynamical viewpoint why the absolute zero temperature ($T=0$ Kelvin(K)) is unattainable.

 Besides the above reasoning, we can also use the idea of thermal capacity $C_T$ of an ideal gas. The third law of thermodynamics states
that $C_T=dQ/dT\rightarrow 0$ in the limit $T\rightarrow 0$K, so that it becomes more and more difficult to withdraw heat from the gas
close to $T=0$K and therefore $0$K becomes unattainable. This phenomenological explanation for the third law of thermodynamics can be
justified by taking into account the new dynamical effects close to $V$. To do that, consider the thermal capacity, namely:

\begin{equation}
C_T= Mc_s,
\end{equation}
where $M$ is the total mass of the gas and $c_s$ is its specific heat for constant volume, being $M=Nm$, and $m(=m_0\Psi)$ is the
 relativistic (bare) mass of each ``particle" (atom or molecule) of the gas. As we intend to introduce the dynamical effects only close to
 $V$, we have the approximation $M=Nm\approx Nm_0\theta(\left<v^2\right>)$. Thus we write:

\begin{equation}
C_T=\frac{dQ}{dT}=Nmc_s\approx N\left(m_0\sqrt{1-\frac{V^2}{\left<v^2\right>}}\right)c_s,
\end{equation}
where $m\approx m_0\theta(\left<v^2\right>)=m_0\sqrt{1-\frac{V^2}{\left<v^2\right>}}$. Here in this regime we find $\Psi(\left<v^2\right>)\approx
 \theta(\left<v^2\right>)$.

From (52) we see that $M\rightarrow 0$ when $\sqrt{\left<v^2\right>}\rightarrow V$, which leads to $C_T\rightarrow 0$. However, since $V$
 is an unattainable inferior barrier, the thermal capacity of the gas will never vanish and $T=0K$ will never be attained. That is the
 fundamental connection between the macroscopic (phenomenological) description of the third law and the new microscopic dynamics of each
 ``particle" governed by the energy barrier of the mimimum speed $V$.

\subsection{The overlap of wave-functions in a condensate}

According to the de-Broglie equation in SSR (eq.45), for low velocities we get the following approximation:

\begin{equation}
\lambda\approx\frac{h}{m_0v\sqrt{1-\frac{V^2}{v^2}}},
\end{equation}
where $v$ is the velocity of a single particle. However, since we have a gas with a very large number $N$ of identical ``particles" like atoms or
molecules, the mean value of wavelength per ``particle"  is

\begin{equation}
\left<\lambda\right>\approx\frac{h}{m_0\sqrt{\left<v^2\right>}\sqrt{1-\frac{V^2}{\left<v^2\right>}}}=
\frac{h}{m_0\sqrt{\left<v^2\right>- V^2}}
\end{equation}

In the newtonian approximation, where $\sqrt{\left<v^2\right>}>>V$, we have higher temperatures. So by making such an approximation in
 (54), we find $\left<\lambda\right>\approx h/m_0\sqrt{\left<v^2\right>}$. And according to (47), we get
 $m_0\sqrt{\left<v^2\right>}=\sqrt{3k_Bm_0T}$ (with $\nu=3$), so that the newtonian approximation can be written as follows:

\begin{equation}
\left<\lambda\right>\approx\frac{h}{\sqrt{3k_Bm_0T}},
\end{equation}
where we have $\left<\lambda\right>\sim T^{-1/2}$ (newtonian regime). But, now if we are interested in the regime close to $V$ (or
close to $T=0$K), we should combine the relation (54) with (49) such that we find $\left<\lambda\right>\sim g(T)$, where $g(T)$ is a
function that recovers the classical regime ($T^{-1/2}$) for higher temperatures.

Our next step is to obtanning $g(T)$. In order to do that, first of all we solve the equation (49) and so we get the physical solution,
as follows:

\begin{equation}
\left<v^2\right>-V^2=\frac{V^2}{2}\left[\sqrt{1+\left(\frac{6fk_BT}{m_0V^2}\right)^2}-1\right],
\end{equation}
so that, if the temperature $T\rightarrow 0$K, this implies $\left<v^2\right>\rightarrow V^2$ or 
$\sqrt{\left<v^2\right>}\rightarrow V$.

 Now, introducing (56) into (54), we finally obtain

\begin{equation}
\left<\lambda\right>=\frac{h}{m_0V}g(T),
\end{equation}
where

\begin{equation}
g(T)=\frac{\sqrt{2}}{\sqrt{\sqrt{1+\left(\frac{6fk_BT}{m_0V^2}\right)^2}-1}}
\end{equation}

Here it is important to call attention to the fact that the wavelength $h/m_0V$ that appears in (57) represents a universal constant
depending on mass $m_0$. For instance, if we consider $m_0$ being the electron mass $m_e$, we find a macroscopic universal wavelength
$\lambda_e=h/m_eV$, which is much larger than the Compton wavelength $\lambda_C$, i.e., $\lambda_e(=h/m_eV)>>\lambda_C(=h/m_ec)$, since
$V<<c$. In another work, we have estimated the minimum speed $V\sim 10^{-15}$m/s\cite{2}. So we get $\lambda_e=h/m_eV\sim 10^{12}$m,
which has practically the inverse of the magnitude of $\lambda_C$. For the case of Hydrogen gas that can form a condensate, we find
$\left<\lambda\right>=\lambda_Hg(T)$, where $\lambda_H=h/m_HV\sim 10^{9}$m, as we have $m_H\sim 10^{3}m_e$. And so on for
any atomic or molecular gas. Thus we realize that $\left<\lambda\right>=h/m_0V$ only for $g(T)=1$ (see (57)), and so there should be a
specific temperature $T_0$ for each gas such that our condition ($g(T_0)=1$) is satisfied. So, from (58) we find 
$T_0=\sqrt{2}m_0V^2/3fk_B$ in order to get $\left<\lambda\right>=h/m_0V$.

We observe that, when $T\rightarrow 0$K, we find $\sqrt{\left<v^2\right>}\rightarrow V$ and $g(T)\rightarrow\infty$ according to (58),
 leading to
 $\left<\lambda\right>\rightarrow\infty$ in (57). So we
have a drastic enlargement of the wavelengths of the ``particles" inside the box, so that they overlap themselves by losing their identities to become
effectively a single huge ``particle" like a super-atom (or super-molecule). Such a huge ``particle"  occupies the entire space inside the box so that it effectively loses its degrees of freedom $\nu$, i.e., $\nu_{eff}\approx 0$.

From (58), for higher temperatures ($k_BT>>m_0V^2$ and $f\approx 1$), we find the function $g(T)\sim 1/\sqrt{T}$ (classical regime) and so
the relation (57) recovers exactly the relation (55).

 In the classical case (higher temperatures), a punctual particle moving inside the box has freedom $\nu=D=3$, however the super-particle,
 that is almost stationary inside the box (very low temperatures: $v$ close to $V$), does not present degrees of freedom
($\nu_{eff}\approx 0$)
 since we get $lim_{T\rightarrow 0}f(T)=0$. This non-classical result will be shown in a coming work, where we will intend to show that the
 function $f$ is of the form $f(T)\approx exp[-(m_0V^2/k_BT)^2]$ so that, just for $T>>m_0V^2/k_B$, we recover the classical case, i.e.,
 $f(T)\approx 1$. For $T\leq m_0V^2/k_B$, corrections are needed. Such a result will be shown elsewhere.

 In this section, we have elaborated a thermodynamic model for a gas considering the existence of a new invariant scale connected to the
 absolute zero temperature, namely a minimum speed that provides the lowest limit for the speed range of a particle; and the speed
 of light in vacuum that sets the upper limit for the speed range of a particle. As Physics is a science based on experimental method, in
 this context all modeling of reality must pass by the scrutiny of the trial. With regard to the model presented here, it is known to
the author that there is no experimental results to be compared with the proposed model. However this fact is not a disappointment since
 many contributions to the advancement of science emerged initially as theoretical models in a search for experimental verifications.
Thus it is expected that the model presented in this paper will stimulate researchers to conduct experiments to verify the results.

\section{\label{sec:level1} Conclusions and prospects}

 We have introduced a space-time with symmetry so that the range of velocities is $V<v\leq c$, where $V$ is an inferior and unattainable
 limit of speed
associated with a privileged inertial reference frame of universal background field (ultra-referential $S_V$). There is a possible connection between the
 minimum speed ($V$) and the minimum length $l_P$ (Planck scale) to be investigated in a further work\cite{2}. The origin of $V$ should be
 also investigated\cite{2}. Actually we will show that $V$ arises from an extension of gravity coupled to the electromagnetic field for
 large distances,
 which could form a basis for understanding a new quantum gravity at very low energies. So we will intend to estimate the scale of $V$ and
 its dependence with $G$, $\hbar$ and some other universal constants\cite{2}. Besides this, within non-commutative geometry and quantum
 deformed Poincare symmetries, we will look for a new kind of geometry and deformed Poincare group that includes the
 minimum speed we are proposing in SSR\cite{3}.

Our relevant investigation was with respect to the problem of the absolute zero temperature in the thermodynamics of an ideal gas. We have made a connection
 between the 3rd law of Thermodynamics and the new dynamics of SSR by means of a relation between the absolute zero temperature ($T=0K$)
 and the minimum average
 speed ($\left<v\right>_N=V$) for a gas with $N$ particles (molecules or atoms). Since $T=0K$ is thermodynamically unattainable, we have
 shown this is due to
 the impossibility of reaching $\left<v\right>_N=V$ from the new dynamics standpoint. This leads yet to another important implication to be
 treated
in detail elsewhere, such as the Einstein-Bose condensate and the problem of the high refraction index of ultracold gases, where we will
intend to estimate that the speed of light would be close to $V$ inside the condensate medium when $T\rightarrow 0K$ and so check
our result against low temperature experiments.

  We will make a more detailed development of the physical consequences of SSR in terms of field-theory actions and gravitational
 extensions.

 The present theory has also various other implications which shall be investigated in the coming articles. We should investigate the
 general transformations of velocity and whether new transformations in SSR can form a group. We will propose a detailed development of a
 new relativistic dynamics where the energy
 of vacuum (ultra-referential $S_V$) plays a crucial role for understanding the origin of the inertia, including the problem of mass
 anisotropy. A new relativistic electrodynamics in the presence of $S_V$ shall be also developed.

 Here we must stress that the covariance of the Maxwell wave equations by change of reference frames in the presence of the background
 field of the ultra-referential $S_V$ has been verified in a previous publication\cite{3}.

  In short we hope to open up a new fundamental research field for various areas of Physics, since the minimum speed can help us to clarify
 several physical concepts, including problems in condensed matter, quantum field theories, cosmology (dark energy and cosmological
 constant\cite{3}) and specially a new exploration for quantum gravity at very low energies (very large wavelengths).\\


\begin{thebibliography}{99}

\bibitem{1} R. Bluhm,arXiv:hep-ph/0506054;S.M.Carroll,G.B.Field and R.Jackiw,
Phys.Rev.D {\bf 41},1231 (1990).
\bibitem{2} C. Nassif: arXiv:gr-qc/1308.5258. 
\bibitem{3} C. Nassif, Pramana-J.Phys., Vol.{\bf 71}, n.1, p.1-13 (2008).
\bibitem{4} D. W. Sciama, {\it On the origin of inertia}. Monthly Notices of the
Royal Astronomical Society {\bf 113}, 34-42 (1953).
\bibitem{5} E. Schr\"{o}dinger,{\it Die erfullbarkeit der relativit\"atsforderung
 in der klassischem mechanik}. Annalen der Physik {\bf 77}, 325-336 (1925).
\bibitem{6} E. Mach,{\it The Science of Mechanics - A Critical and
Historical Account of Its Development \/}. Open Court, La Salle, 1960.
\bibitem{7} A. Einstein, {\it\"{A}ether und Relativit\"{a}ts-theory}, Berlin:
J.Springer Verlag,1920;C.Eling,T.Jacobson,D.Mattingly, arXiv:gr-qc/0410001.
\bibitem{8} Letter from A. Einstein to H. A. Lorentz, 17 June 1916; {\bf item 16-453} in Mudd Library, Princeton University.
\bibitem{9} A. Einstein and L. Infeld, {\it ``The Evolution of Physics'' \/}, Univ. Press Cambridge, 1947.
\bibitem{10} P. Jordan, {\it ``Albert Einstein''}, Verlag Huber, Frauenfeld und Stuttgart, 1969.
\bibitem{11} A. Einstein, {\it Schweiz. Naturforsch. Gesellsch.,Verhandl.} {\bf 105}, 85-93. (1924).
\bibitem{12} A. Einstein, {\it \'Ather und Relativit\"as-theorie \/}, Springer, Berlin, 1920. See also: A. Einstein, Science, {\bf 71}, 608-610 (1930); Nature,{\bf 125}, 897-898.
\bibitem{13} I. Licata, Hadronic J.{\bf 14}, 225-250 (1991).
\bibitem{14} G. A. Camelia, Int. J. Mod. Phys. D{\bf 11}, 35-60, (2002).
\bibitem{15} G. A. Camelia et al., Nature {\bf 393}, 763-765 (1998).
\bibitem{16} G. A. Camelia, Nature {\bf 418} 34 (2002).
\bibitem{17} G. A. Camelia, Int. J. Mod. Phys. D{\bf 11}, 1643 (2002).
\bibitem{18} J. Magueijo and L. Smolin, Phys. Rev. Lett.{\bf 88}, 190403 (2002).
\bibitem{19} J. Magueijo and A. Albrecht, Phys. Rev. D{\bf 59}, 043516 (1999).
\bibitem{20} J. Magueijo and L. Smolin, Phys. Rev. D{\bf 67}, 044017 (2003).
\bibitem{21} J. K. Glikman and L. Smolin, Phys. Rev. D{\bf 70}, 065020 (2004). See in: arXiv:hep-th/0406276.
\bibitem{22} F. Girelli and E. R. Livine, Phys. Rev. D{\bf 69}, 104024 (2004). See in: arXiv:gr-qc/0311032.
\bibitem{23} H. F. M. Goenner, {\it On the History of Geometrization of Space-time}, arXiv:gr-qc/0811.4529v1 (2008).
\bibitem{24} H. Busemann, {\it The Geometry of Finsler Space}. Bulletin of the American Mathematical Society 56, 5-16 (1950).
\bibitem{25} E. Cartan, {\it Les Espaces de Finsler}. Actualit\`es Scientifiques et Industrielles, No. 79, Paris: Hermann (1934).
\bibitem{26} P. Finsler, {\it Uber Kurven und Fl\"{a}chen in allgemeinen R\"{a}umen}. Diss. Univ. G\"{o}ttingen 1918;
Nachdruck mit zus\"{a}tzlichem Literaturverzeichnis. Basel: Birkh\"{a}user (1951).
\bibitem{27} F. Girelli, S. Liberati and L. Sindoni, Phys. Rev.{\bf D 75}, 064015 (2007).
\bibitem{28} C. J. Adler, Am. J. Phys.{\bf 55}, 739 (1987).
\bibitem{29} R. P. Feynman, R.B. Leighton and M. Sands, {\it Th Feynman Lectures on Physics},
 Addison-Wesley Reading, MA (1963), Vol. 1, sections 15-8 and 16-4.
\bibitem{30} L. V. Okun, Physics Today {\bf 42}(6), 31 (1989).
\bibitem{31} T. R. Sandin, Am. J. Phys. {\bf 59}, 1032 (1991).
\bibitem{32} W. Rindler, {\it Introduction to Special Relativity}, Clarendon Press, Oxford (1982), pp. 79-80;
 W. Rindler,{\it Essential Relativity}, Springer, New York (1977), 2nd. edition, Section 5.3.
\bibitem{33} E. F. Taylor and J. A. Wheeler, {\it Spacetime Physics}, W. H. Freeman Co., New York (1992), 2nd. edition, pp. 246-251.
\end{thebibliography}
\end{document}